\documentclass[twocolumn,showpacs,preprintnumbers,a4,prl]{revtex4}
\usepackage{graphicx}%
\usepackage{dcolumn}
\usepackage{amsmath}
\usepackage{amssymb}
\usepackage{bm}

\begin{document}

\title{From the second magnetization peak to peak effect.
A study of superconducting properties in Nb films and MgB$_2$ bulk
samples}
\author{Dimosthenis Stamopoulos,\footnote[3]{Corresponding author
(densta@ims.demokritos.gr)} Athanasios Speliotis and Dimitris
Niarchos}

\affiliation{Institute of Materials Science, NCSR "Demokritos",
153-10, Aghia Paraskevi, Athens, Greece.}
\date{\today}

\begin{abstract}
We report on magnetic and magnetoresistance measurements in two
categories of superconducting Nb films grown via magnetron
sputtering and MgB$-2$ bulk samples. In the first category, films
of $T_c=9.25$ K were produced by annealing during deposition. In
these films, the magnetic measurements exhibited the so-called
''second magnetization peak'' (''SMP''), which is accompanied by
thermomagnetic instabilities (TMI). The characteristic field
$H_{\rm fj}$, where the first flux jump occurs, has been studied
as a function of the sweep rate of the magnetic field.
Interestingly, in the regime $T<6.4$ K, the respective line
$H_{\rm fj}(T)$ is constant, $H_{\rm fj}($T$<6.4$ K)$=40$ Oe. A
comparison to TMI observed in MgB$_2$ bulk samples is also
performed. Our experimental findings can't be described accurately
by current theories on TMI. In the second category, films of
$T_c=8.3$ K were produced without annealing during deposition. In
such films, we observed a peak effect (PE). In high magnetic
fields the PE is accompanied by a sharp drop and a narrow
hysteretic behavior ($\Delta T< 20$ mK) in the measured
magnetoresistance. In contrast to experimental works presented in
the past, the comparison of our magnetic measurements with the
magnetoresistance data suggests that rather the appearance of
surface superconductivity than the melting transition of vortex
matter, is the cause of the observed behavior.

\end{abstract}


\maketitle

\section{Introduction}

The discovery of compounds that exhibit high critical temperatures
opened a new wide field in the physics of superconductivity
\cite{blatter94}. A consequence of the ongoing theoretical and
experimental interest for the effects observed in high-$T_c$
compounds, is the reexamination of the related topics in the more
isotropic low-$T_c$ superconductors
\cite{Mikitik01,Menon02,Sarkar01}. Interestingly, current
theoretical studies try to unify the phase diagram of vortex
matter by appropriately taking into account all the factors that
influence the behavior of the magnetic flux lines created in both
low and high-temperature superconductors \cite{Mikitik01,Menon02}.

Isotropic Nb is a low-$T_c$ conventional superconductor which
recently has become a subject of intensive experimental studies
\cite{Drulis91,Schmidt93,Lynn94,Salem02,Gammel98,Ling01,Forgan02,Brandt02,Ling02}.
One of the unresolved issues is the nature of the vortex state
which is settled in the regime close to the upper-critical field
$H_{\rm c2}(T)$. More specifically, questions on the importance of
fluctuation effects \cite{Salem02} and the nature of the melting
transition and/or an order-disorder transition is still under
investigation in this type-II superconductor
\cite{Gammel98,Ling01,Forgan02,Brandt02,Ling02}. In addition,
recent magnetic studies in Nb films
\cite{Esquinazi99,Kopelevich98} reported the existence of a
structure that reminisces of the second magnetization peak (SMP)
which is usually observed in high-$T_c$ superconductors
\cite{Zhukov95,Stamopoulos01,Stamopoulos02,Stamopoulos03,Sun00}.
These studies concluded that for the case of Nb films the ''second
magnetization peak'' (''SMP'') is motivated by thermomagnetic
instabilities (TMI) that occur in the low-temperature regime, far
below the upper-critical field
\cite{Nowak97,Esquinazi99,Kopelevich98}. Despite the need for the
complete theoretical understanding of the underlying mechanism
that motivates the ''SMP'', the existence of the accompanying TMI
should be studied experimentally in more detail, because the
undesirable flux jumps constitute a serious limitation for
practical applications. Finally, in the last years many
experimental and theoretical works dealt with the change of the
superconducting properties of Nb films, when placed in close
proximity with arrays of ferromagnetic particles or magnetic
homogenous layers \cite{Lange03,Martin99,Velez02a,StamopoulosHS}.
Such composite structures may constitute a starting point for the
generation of important electronic devices in the near future
\cite{Lange03}. Thus, thorough studies of the phase diagram of
vortex matter in pure Nb, and other low-$T_c$ superconducting
films, may give information that could be important in other areas
of science.

In spite of the ongoing experimental research, a complete study of
the phase diagram of vortex matter in an extended
temperature-magnetic-field regime is still lacking, leaving the
above mentioned issues still open. To this end, we performed
systematic magnetic and transport measurements in sputtered
polycrystalline films of the low-$T_c$ Nb superconductor. In
relatively thick samples, {\it produced by annealing during the
deposition}, the magnetic measurements exhibited a ''SMP'' at
points $H_{\rm ''smp''}(T)$, which is placed well below the
upper-critical fields $H_{\rm c2}(T)$. The ''SMP'' is accompanied
by flux jumps, occurring for $H<H_{\rm ''smp''}(T)$, while for
$H>H_{\rm ''smp''}(T)$ smooth m(H) curves are observed. For low
enough temperatures $T<6.4$ K, the first flux jump line $H_{\rm
fj}(T)$, where the first flux jump in the virgin magnetic curve is
observed, becomes constant $H_{\rm fj}($T$<6.4$ K)$=40$ Oe.
Furthermore, the characteristic field $H_{\rm fj}(T)$ is not
inversely proportional to the sweep rate of the applied field.
These experimental findings are in strong contrast to theoretical
expectations and remain to be explained. The first flux jump line
$H_{\rm fj}(T)$ ends at a characteristic temperature $T_o=7.2$ K,
where it connects with the first peak $H_{\rm fp}(T)$, and with
the ''SMP'' $H_{\rm ''smp''}(T)$ lines. For $T>T_o=7.2$ K no flux
jumps were observed, in nice agreement to theoretical suggestions.
In order to compare the TMI observed in films and in bulk samples
we also present supplementary magnetic data in a MgB$_2$ bulk
sample. The comparison revealed many differences in the observed
behaviors.

In thinner films, {\it produced without annealing during the
deposition}, transport measurements revealed a conventional peak
effect (PE), placed in the vicinity of the upper-critical fields
$H_{\rm c2}(T)$. For high enough magnetic fields, the response at
the end points $H_{\rm e\
 p}(T)$ of the PE exhibits a narrow hysteretic behavior ($\Delta T< 20$ mK).
The hysteresis faints as we move in the low-field regime. In
addition, the hysteretic response is current dependent (not
observed in very small or high currents), indicating that is
motivated by a dynamic cause. A discussion referring to surface
superconductivity and to the melting and/or disordering transition
of vortex matter in Nb is made.

This paper is organized as follows. In Sec. II we give details
about the preparation of the films and the performed experiments.
Section III presents magnetic and transport data for the film
Nb$-1$ that exhibits the ''SMP'' and the TMI. The vortex matter
phase diagram is constructed, and a discussion is made in
comparison to current theories on TMI. A brief comparison with
recent magneto-optical studies in Nb and MgB$_2$ films is also
made. In Sec. IV we present experimental results for the film
Nb$-2$ that exhibits the PE. The phase diagram of vortex matter
for the Nb$-2$ sample is also presented. An extended discussion on
the disordering and the melting transitions, and also on surface
superconductivity is made. In Sec. V we present crystallographic
results coming from transmission electron microscopy (TEM) and
x-ray diffraction (XRD) experiments and we make a comparative
discussion on the results that the two different films exhibit.
Finally, the conclusions are presented in Sec. VI.

\section{Preparation of the films and Experimental details}

The samples of Nb were sputtered on Si $[001]$ substrates under an
Ar atmosphere ($99.999$ \% pure). The base pressure was $5\times
10^{-7}$ Torr. We examined in detail the influence of the
deposition rate and of the annealing temperature (during the
deposition) on the quality of the produced films. As a criterion,
we measured the upper-critical field line $H_{\rm c2}(T)={\Phi
_o/2\pi \xi^2(T)}$ of the produced Nb films. The films that were
produced at deposition rates of the order of $4$ {\AA}/sec (at a
DC power of $55$ Watt) exhibited the smallest slope
$dH_{c2}(T)/dT$. In this work we present data on two categories of
sputtered films. The first category refers to films produced by
annealing at $T=300$ C during deposition. Such films exhibit a
critical temperature $9.25$ K, equal to the one of high purity
single crystals \cite{Finnemore66}. The second category refers to
films produced without annealing during the deposition. These
films showed a critical temperature of $8.3$ K. Below we present
results for a representative Nb film of each category. The Nb film
of the first category is labelled as Nb$-1$ and has $T_c=9.25$ K,
while the one of the second category is labelled as Nb$-2$ and has
$T_c=8.3$ K. The residual resistance ratio was R($300$ K)/R($10$
K)=$6.8$ and $2.9$ for Nb$-1$ and Nb$-2$ respectively. Their
thickness is $7700$ {\AA} and $1600$ {\AA} for Nb$-1$ and Nb$-2$
respectively.

Our magnetic measurements were performed by means of a commercial
SQUID magnetometer (Quantum Design). In all magnetic data
presented below, the magnetic field was always normal to the
surface of the film (${\bf H}\parallel{\bf c}$). All the
magnetization data were obtained under zero field cooling (ZFC).
In this experimental protocol, by starting from a temperature
above $T_c$, the sample is cooled to the desired temperature under
zero magnetic field. In this way, we obtain the virgin
magnetization curves. Our magnetoresistance measurements were
performed by applying a dc transport current and measuring the
voltage in the standard four-point configuration. In the transport
data presented in this work, the magnetic field was mainly normal
to the surface of the film (${\bf H}\parallel{\bf c}$). The cases
where the presented data refer to a field parallel to the film's
surface, will be explicitly specified. For both field
orientations, the applied current was normal to the magnetic field
${\bf J}_{\rm dc}\perp{\bf H}_{\rm dc}$, so that the vortex lines
experience a non-zero Lorentz force ${\bf F}_{\rm L}\propto{\bf
J}_{\rm dc}\times{\bf H}_{\rm dc}$. In our magnetoresistance
measurements we also employed the ZFC initial conditions. Thus, a
direct comparison with our magnetic data could be made safely. The
temperature control and the application of the dc fields in our
transport measurements were achieved in our SQUID magnetometer. We
examined the whole temperature-magnetic-field regime accessible by
our SQUID ($H_{\rm dc}<55$ kOe, $T>1.8$ K).

\section{Experimental results and discussion for sample Nb$-1$}

\subsection{Thermomagnetic instabilities and the ''second magnetization peak''}

\begin{figure}[tbp] \centering%
\includegraphics[angle=0,width=8cm]{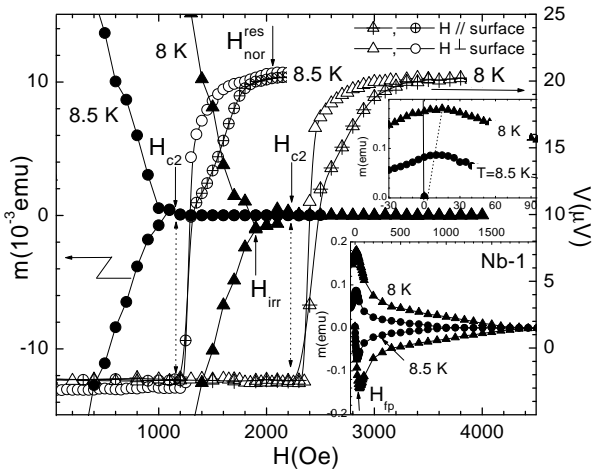}
\caption {Isothermal magnetic m(H) and magnetoresistance V(H)
measurements for the film Nb$-1$ (solid and open points
respectively) in two representative temperatures $T=8$ and $8.5$
K, focused in the regime of the upper-critical field $H_{\rm c2}$.
The solid and open points without (with) crosses refer to the case
where the magnetic field is normal (parallel) to the surface of
the film. In both transport data the applied current ($I_{\rm
dc}=1$ mA) was normal to the magnetic field ${\bf J}_{\rm
dc}\perp{\bf H}_{\rm dc}$. The first inset presents the whole
loops, so that the first peak $H_{\rm fp}$ may be seen, while the
second inset focusses on the descending branch of the loops so
that the respective peak may be observed.}
\label{b5}%
\end{figure}

We start the presentation of our experimental results for sample
Nb$-1$, which in the low-temperature regime exhibits the TMI and
the ''SMP''. Figure \ref{b5} presents magnetic measurements in
comparison to transport data, in the high-temperature regime, for
two representative temperatures $T=8$ and $8.5$ K. In magnetic
measurements the applied field is normal to the surface of the
film (as in all the magnetic data presented in this work), while
in transport data we present both cases where the field is normal
(open points) and parallel (points with crosses) to the surface of
the film.

First of all, we see that, with high accuracy, the measured
voltage becomes non-zero at the magnetically determined
upper-critical points $H_{\rm c2}$ (where the magnetic moment
vanishes). The simultaneous occurrence of the resistive and the
magnetic transitions is expected in conventional low-$T_c$
superconductors \cite{Abrikosov88}. We observe that the comparison
of transport data to magnetic measurements is very informative. A
lack of such a comparison could mistakenly lead to the
interpretation of a melting transition at the points where the
sharp drop in the voltage curves is observed. Our results reveal
that in low-$T_c$ superconductors, sharp resistance drops are also
caused by the conventional transition between the normal and
superconducting states. Above the upper-critical points $H_{\rm
c2}$ we observe a rounding of the measured voltage curve, which
attains the normal state value at a much higher field $H_{\rm
nor}^{\rm res}$. This feature is more evident when the magnetic
field is parallel to the surface of the film (symbols with
crosses), indicating that it is probably related to surface
superconductivity (see below) \cite{Abrikosov88,James63}. In the
lower inset we present the whole magnetic loops so that the first
peak $H_{\rm fp}$ may be seen. The upper inset focusses in the
low-field part of the descending branch so that the respective
peak may be easily seen. We observe that the peak of the
descending branch is placed in positive field values. In a recent
work Shantsev et al. \cite{Shantsev99} have studied the position
of the peak observed in the descending branch in m(H) loops when
the magnetic field is normal to a superconducting specimen. They
concluded that this peak is placed to positive field values when
the sample exhibits a granular structure \cite{Shantsev99}. In our
case TEM data revealed that the Nb$-1$ film exhibits a slight
tendency for columnar growth (see below). Thus, in agreement to
Ref. \onlinecite{Shantsev99} we suggest that the peak of the
descending branch is placed to positive field values due to the
underlying columnar growth of the Nb$-1$ film.

\begin{figure}[tbp] \centering%
\includegraphics[angle=0,width=8cm]{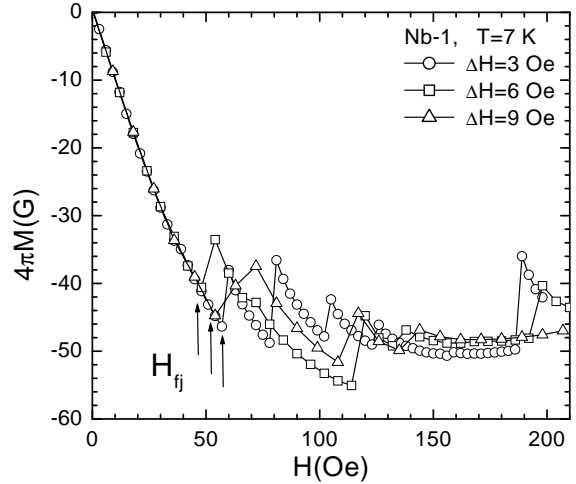}
\caption {Isothermal magnetization curves $4\pi$M(H) for film
Nb$-1$ at $T=7$ K, for three different sweep rates of the magnetic
field with steps $\Delta H=3, 6$ and $9$ Oe. We observe that for
higher sweep rate of the magnetic field, flux jumps are more rare.
In the resulting magnetization curves demagnetization corrections
have been taken into account \cite{dema,Fetter67}.}
\label{b6}%
\end{figure}%

\begin{figure}[tbp] \centering%
\includegraphics[angle=0,width=8cm]{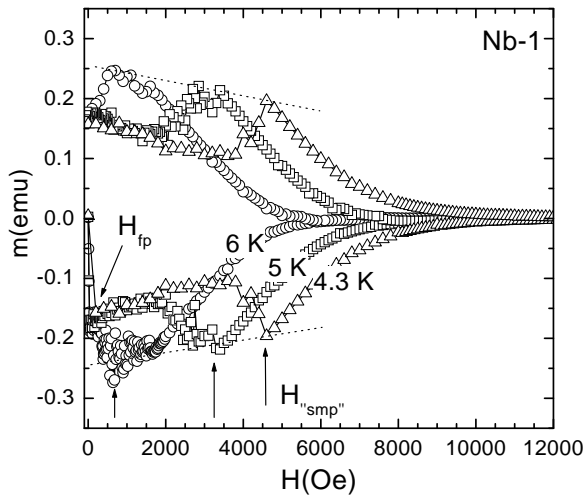}%
\caption { Isothermal magnetic moment m(H) measurements for Nb$-1$
sample, in the low-temperature regime at $T=4.3, 5$ and $6$ K. In
addition to the first peak $H_{\rm fp}$, which is placed in the
low-field regime, we observe a ''SMP'' in higher magnetic fields
$H_{\rm ''smp''}$. The peak value of the ''SMP'' decreases for
lower temperatures (see dot lines which serve as guides to the
eye). }
\label{b7}%
\end{figure}

In fig. \ref{b6} we present magnetization measurements
\cite{dema,Fetter67} for $T=7$ K and for different field sweep
rates, focused in the low-field regime where the first peak
$H_{\rm fp}$ should be expected. In contrast to the smooth curves
which are observed at temperatures close to $T_c$, we see that for
$T=7$ K flux jumps are present. The observed flux jumps are small,
of the order of a few Gauss to less than twenty Gauss. More
importantly, the flux jumps depend on the sweep rate of the field.
As we increase the sweep rate, the flux jumps are rare. At even
lower temperatures the magnetic response becomes more anomalous.
At $T=6$ K a broad ''noisy'' first peak shows up, which at lower
temperatures transforms into two separate peaks: a first maximum
which occurs in low-field values, and a distinct second peak which
is placed in high magnetic fields. This behavior may be seen in
fig. \ref{b7} for three representative temperatures $T=4.3, 5$ and
$6$ K. The overall behavior resembles the SMP observed usually in
high-$T_c$ superconductors
\cite{Zhukov95,Stamopoulos01,Stamopoulos02,Stamopoulos03,Sun00}.
For this reason we ''loosely'' refer to these characteristic
fields as $H_{\rm ''smp''}$. However, there are noticeable
differences between the magnetic behaviors  observed in Nb$-1$ and
in high-$T_c$ superconductors. Below the ''SMP'' ($H<H_{\rm
''smp''}$) the response is ''noisy'', while above it ($H>H_{\rm
''smp''}$) we observed smooth magnetic curves. In contrast, in
{\it point disordered} high-$T_c$ superconductors the
magnetization curves are smooth, both below and above the SMP. In
addition, the peak value of the $H_{\rm ''smp''}$ slightly
decreases as we lower the temperature (see fig. \ref{b7}). In
high-$T_c$ superconductors for lower temperatures the peak value
of the SMP strongly increases \cite{Zhukov95,Stamopoulos01,Sun00}.
Finally, the resulting line $H_{\rm ''smp''}(T)$ ends on the first
peak line $H_{\rm fp}(T)$ at $T_o/T_c\approx 0.78$ in our Nb$-1$
film (see fig. \ref{b12}(b) below), while in high-$T_c$
superconductors the respective line $H_{\rm smp}(T)$ ends near the
irreversibility/melting line, or extends almost up to the critical
temperature.

\begin{figure}[tbp] \centering%
\includegraphics[angle=0,width=8cm]{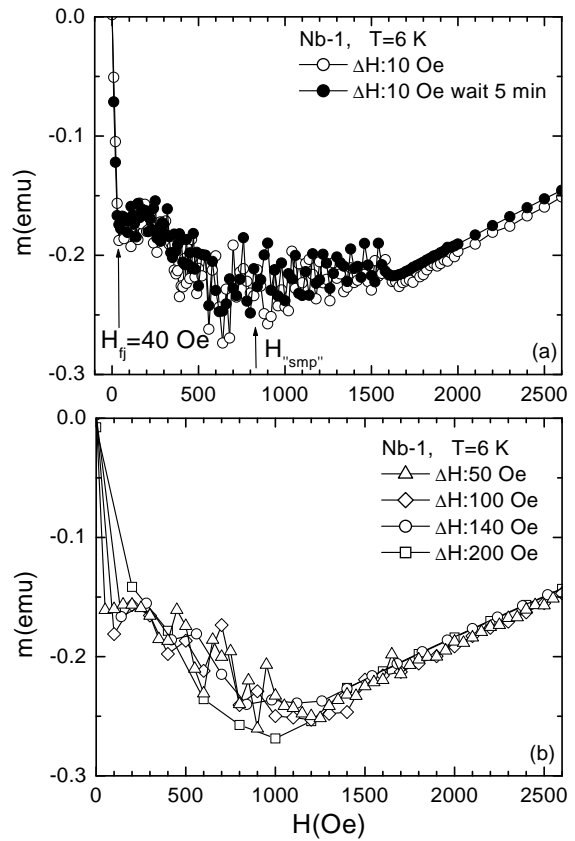}%
\caption { Isothermal magnetic moment m(H) measurements for film
Nb$-1$ at $T=6$ K, for different application rates of the magnetic
field with steps, $\Delta H=10$ Oe (upper panel) and $\Delta H=50,
100, 140$ and $200$ Oe (lower panel). When introducing a high
enough sweep rate of the magnetic field (lower panel), the
measured curves exhibit flux jumps more rarely. For the case where
$\Delta H=10$ Oe (upper panel) we performed exactly the same
measurement when introducing a waiting time of $5$ min before
start measuring (after the application of the new field value). We
see that the magnetic behavior is exactly the same as when
measuring right after the application of the field. }
\label{b8}%
\end{figure}%

In a recent theoretical study Mints and Brandt \cite{Mints96}
considered the special case where TMI occur in superconducting
films when the magnetic field is normal to the film's surface.
They derived the following criterion for the first field value
$H_{\rm fj}$ where flux jumps occur
\begin{equation}
\frac{H_{\rm fj}(T)\mu _0\acute{H}l}{\kappa(T) q^2j_c(T)}
\frac{dj_c(T)}{dT}=1, \label{eq1}
\end{equation}
in where $\acute{H}$ is the sweep rate of the applied field, $l$
is the exponent of the $J-E$ characteristic
$(J(E)=J_c(E/E_0)^{1/l})$, $\kappa(T)$ is the thermal conductivity
and $q$ is a parameter that accounts for the thermal boundary
resistance between the film and the substrate. This model that
deals with an experimentally realizable case, predicts that the
first flux jump field $H_{\rm fj}$ decreases as the sweep rate
$\acute{H}$ of the applied field and the exponent $l$ of the $J-E$
curves increase.

\begin{figure}[tbp] \centering%
\includegraphics[angle=0,width=8cm]{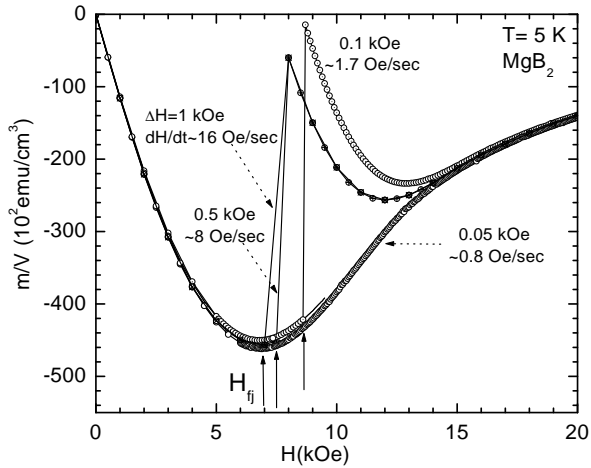}%
\caption { Isothermal magnetic m(H) measurements in a bulk sample
of MgB$_2$ for $T=5$ K, and for various steps of the applied field
$\Delta H=1, 0.5, 0.1$ and $0.05$ kOe. The first flux jump field
$H_{\rm fj}$ is placed to lower values when increasing the sweep
rate of the applied field $\acute{H}$. }
\label{extra}%
\end{figure}%

In order to test the theoretical predictions we performed
systematic magnetic measurements for various sweep rates of the
applied field. In fig. \ref{b8} we present such data at
temperature $T=6$ K. We observe that as we increase the sweep rate
of the applied field, the flux jumps are rare. This indicates that
the flux jumps can not be detected in high electric-field levels
(where vortices are probably unpinned) but are present only when
we probe the vortex system in very low electric-fields.
Furthermore, we studied in detail the dependence of the first flux
jump field $H_{\rm fj}$ on the sweep rate of the magnetic field
for steps $\Delta H$ low enough, so that $H_{\rm fj}$ could be
defined accurately. In measurements (not shown here) performed
with steps $\Delta H$ in the range $2$ Oe$<\Delta H<15$ Oe, we
observed that the field $H_{\rm fj}$ remained constant, $H_{\rm
fj}=40$ Oe. This is in contrast to theoretical expectations. Our
results are in agreement to the experimental studies of Ref.
\onlinecite{Esquinazi99}. On the other hand, such anomalous flux
jumps could be an artifact, caused by the inability of our SQUID
to maintain a constant temperature on the whole film. To test this
case we performed exactly the same measurement for step $\Delta
H=10$ Oe, after having waited for $5$ min before measuring. The
result is presented in fig. \ref{b8}(a). We clearly see that the
flux jumps are still present.

Very recently it was discovered that the compound MgB$_2$ is a
superconductor with relatively high critical temperature $T_c=39$
K. As a result of the renewed interest, many groups have done
reliable studies on TMI in MgB$_2$ samples, either in thin film or
in bulk form, by performing thorough magnetic measurements, or by
using more direct techniques such as magneto-optical imaging
\cite{Barkov03,Johansen02,Johansen01,Zhao02}. The observed
behavior in MgB$_2$ films is very similar to the behavior that we
observed in our Nb films. Thus, in a next section we compare our
experimental results to recent results obtained in MgB$_2$ films.
On the other hand, currently available theories distinguish the
driving mechanisms of TMI when observed in thin films or in bulk
samples. In order to directly compare the two cases, we also
present magnetic measurements performed in a bulk sample of
MgB$_2$ superconductor \cite{PissasStamopoulos}. The preparation
process is reported elsewhere \cite{Pissas01}. In fig. \ref{extra}
we observe that by decreasing the sweep rate of the field, the
first flux jump field $H_{\rm fj}$ increases. This is in agreement
to theoretical expectations for bulk samples
\cite{Wipf6791,Mints81}. In contrast, in the Nb$-1$ film the field
$H_{\rm fj}$ doesn't exhibit such behavior, as already discussed
above. In our bulk MgB$_2$ sample, TMI are not observed for low
sweep rates of the applied field, $\acute{H}<1$ Oe/sec. On the
other hand, for $\acute{H}>1$ Oe/sec, flux jumps are observed that
are comparable to the measured magnetic moment. Thus, after every
flux jump the sample almost enters the normal state. We may then
refer to a global thermal runaway as the underlying mechanism of
TMI in this bulk sample. In contrast, the flux jumps observed in
our Nb$-1$ film are much smaller than its magnetic moment and
don't turn the sample in the normal state. Finally, in our MgB$_2$
bulk sample, the first flux jump field $H_{\rm fj}$ is placed
above the full penetration field, which may be fairly approximated
by the first peak field $H_{\rm fp}$. This is in contrast to
theoretical expectations \cite{Wipf6791,Mints81}. In the Nb$-1$
film, we observed that when the first flux jump field $H_{\rm fj}$
exceeds the first peak field $H_{\rm fp}$, TMI are no longer
present (see figs.\ref{b12} and \ref{b13} below). The differences
mentioned above indicate that the mechanism of TMI is not only
quantitatively, but could be also qualitatively different between
bulk and film samples.

\begin{figure}[tbp] \centering%
\includegraphics[angle=0,width=8cm]{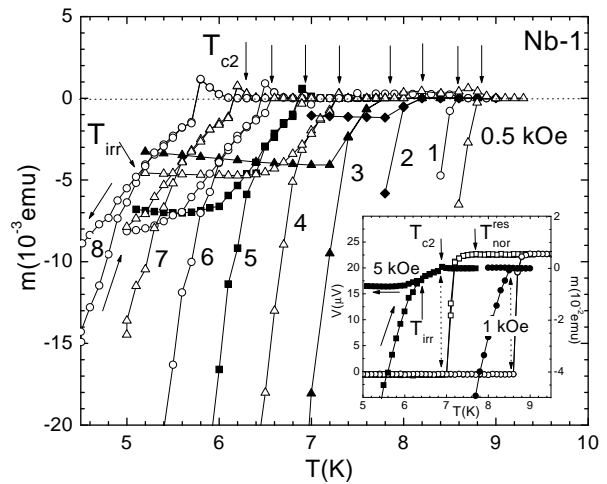}%
\caption { Isofield magnetic measurements m(T) for Nb$-1$ under
various magnetic fields $H=0.5-8$ kOe. We focus in the regime near
the upper-critical points $T_{\rm c2}$ so that the irreversibility
points $T_{\rm irr}$ may be seen. The inset presents,
comparatively, magnetic and voltage measurements as a function of
temperature, at fields $H=1$ and $5$ kOe. We observe the
congruence of the magnetic and the resistive transitions (see dot
arrow lines). At the irreversibility points $T_{\rm irr}$ the
voltage does not show any measurable feature. }
\label{b10}%
\end{figure}%

\subsection{Phase diagram of vortex matter for Nb$-1$. Comparison with theory}

In order to construct the phase diagram of vortex matter for
Nb$-1$ film, we also performed isofield magnetic measurements as a
function of temperature. Representative data are shown in fig.
\ref{b10}. We mainly focus near the transition to the normal
state, so that the irreversibility points $T_{irr}$ may be easily
seen. The inset presents comparative magnetic moment and voltage
measurements as a function of temperature. We observe that as we
increase the applied field, the regime of magnetic reversibility
is enhanced. In addition, the magnetic and the resistive
transitions clearly coincide, in agreement to our isothermal
measurements as a function of field (see fig. \ref{b5}). At the
irreversibility points $T_{irr}$ the measured voltage curves do
not show any distinct feature.

\begin{figure}[tbp] \centering%
\includegraphics[angle=0,width=8cm]{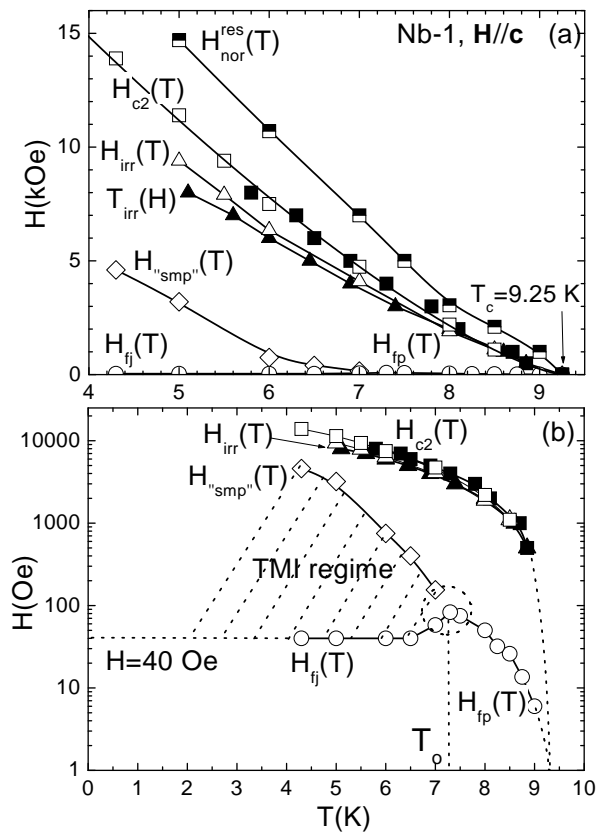}%
\caption {Resulted phase diagram for the Nb$-1$ film (${\bf
H}\parallel{\bf c}$). Presented are the $H_{\rm fp}(T)$ and
$H_{\rm fj}(T)$ lines (low-field open circles), the ''SMP'' line
$H_{\rm ''smp''}(T)$ (rhombi), the $H_{\rm irr}(T)$ (triangles)
and the upper-critical field $H_{\rm c2}(T)$ (squares) lines
(where open points come from isothermal m(H) measurements, while
solid points come from isofield m(T) measurements). The solid line
through the $H_{\rm c2}(T)$ points, is a least squares fitting by
using the expression $H_{\rm c2}(T)=H_{\rm c2}(0)(1-T/T_c)^m$ with
$H_{\rm c2}(0)=31.8\pm 0.8$ kOe and $m=1.34\pm 0.03$. In the upper
panel we also included the characteristic field line $H_{\rm
nor}^{\rm res}(T)$ where the voltage takes the normal state value.
The lower panel reveals the details of the features observed in
the low-field regime in semi-logarithmic scale.}
\label{b12}%
\end{figure}

The resulted ''phase diagram'' for the film Nb$-1$ is presented in
figs. \ref{b12}(a) and \ref{b12}(b). Open (solid) triangles,
coming from magnetic measurements as a function of field
(temperature), denote the irreversibility fields $H_{\rm irr}(T)$
(temperatures $T_{\rm irr}(H)$), while open and solid squares
(from magnetic measurements as a function of field and
temperature, respectively) refer to the magnetically determined
upper-critical fields $H_{\rm c2}(T)$ and temperatures $T_{\rm
c2}(H)$. Furthermore, the semi-filled squares refer to the
characteristic line $H_{\rm nor}^{\rm res}(T)$ where the
resistance takes the normal state value. In the low-field regime,
presented in detail in fig. \ref{b12}(b), the rhombi originate
from the ''SMP'', while the open circles refer to the first peak
field $H_{\rm fp}(T)$ (for $T>T_o=7.2$ K) and to the first flux
jump field $H_{\rm fj}(T)$ (for $T<T_o=7.2$ K).

\begin{figure}[tbp] \centering
\includegraphics[angle=0,width=8cm]{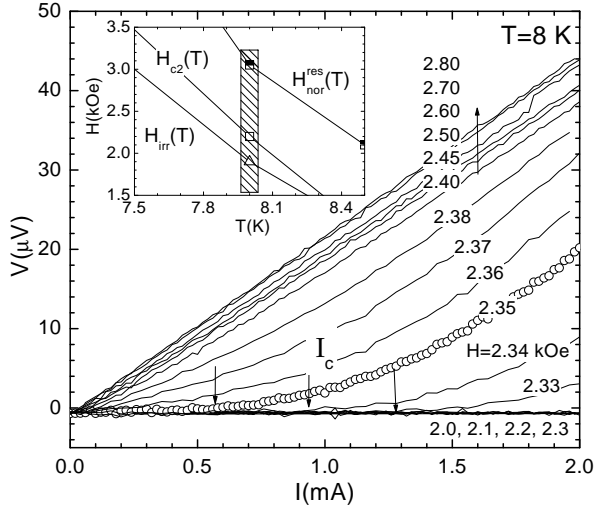}
\caption {Detailed measurements of the $I-V$ characteristics at
$T=8$ K and for various magnetic fields $2$ kOe$<H<2.8$ kOe (main
panel) and a detail of the phase diagram where the measurements
have been performed (inset). Despite the fact that some of the
transport measurements were performed inside the magnetically
reversible regime a linear behavior of the $I-V$ curves (expected
in a liquid state of flux lines) was not observed. The Ohmic
behavior is recovered gradually in the regime $H_{\rm c2}(8\rm
K)<H<H_{\rm nor}^{\rm res}(8\rm K)$.}
\label{b11}%
\end{figure}

First of all, we observe that the $H_{\rm irr}(T)$ data exhibit a
slight upward curvature, which is reminiscent of the behavior
observed in type-II high and low-$T_c$
\cite{Sarkar01,Drulis91,Schmidt93,Stamopoulos01,Stamopoulos03}, or
even type-I \cite{Grover91} superconductors. As we clearly see,
the irreversibility points $T_{\rm irr}(H)$ as determined from
isofield magnetic measurements as a function of temperature (solid
triangles), do not coincide with the respective points $H_{\rm
irr}(T)$ as determined from isothermal magnetic loop measurements
(open triangles). This is a consequence of their dynamic origin.
The irreversibility points, as determined from such measurements,
simply mark the boundary where the vortex system is at a
pseudo-equilibrium state for the specific measuring time of every
experiment. Thus, the irreversibility points ($H_{\rm irr}(T)$ or
$T_{\rm irr}(H)$) should not be attributed to the points where a
melting transition of vortex matter takes place, as in the past
has been reported for the case of sputtered Nb films
\cite{Schmidt93}. Detailed transport measurements of the $I-V$
characteristics that we performed give additional evidence to this
point of view. Representative data are shown in fig. \ref{b11} for
$T=8$ K and various magnetic fields in the regime close to the
upper-critical field $H_{\rm c2}(8\rm K)$. In the inset we present
a detail of the respective part of the phase diagram where the
measurements have been performed (shaded area). We clearly see
that even in the regime above the magnetically determined
irreversibility field $H_{\rm irr}(8\rm K)$ the $I-V$
characteristics are non-linear and gradually attain an almost
linear behavior as the upper-critical field $H_{\rm c2}(8\rm K)$
is approached. This means that in low-T$_c$ superconductors the
irreversibility points can not be ascribed to a true melting
transition since a true liquid state should be accompanied by
absolutely linear behavior in the $I-V$ curves. Furthermore, even
above the upper-critical field a slight nonlinearity is maintained
in the $I-V$ curves which is removed only above the characteristic
field $H_{\rm nor}^{\rm res}(8\rm K)$ where the voltage attains
its normal state value. This fact indicates that the regime
$H_{\rm c2}(T)<H<H_{\rm nor}^{\rm res}(T)$ of the phase diagram
refers to surface superconductivity (see below)
\cite{James63,Abrikosov88}. Finally, we note that the present data
suggest that the Nb$-1$ film is quite disordered since even at
$H=2.3$ kOe$\simeq H_{\rm c2}^{\rm mag}(8\rm K)$ the maximum
applied current $I_{\rm dc}=2$ mA is not able to depin vortices.
By taking into account the dimensions of the specific film we
estimate the effective current density which for the case under
discussion ($T=8$ K, $H=2.3$ kOe) is $J_{\rm dc}\simeq 0.1$
kA/cm$^{2}$. Thus, at $T=8$ K, $H=2.3$ kOe the critical current
density $J_{\rm c}$ exceeds $0.1$ kA/cm$^{2}$.

Let us now return to the discussion of the phase diagram presented
in figs.\ref{b12}(a) and \ref{b12}(b). We see that in contrast to
the irreversibility points $H_{\rm irr}(T)$ the upper-critical
points $H_{\rm c2}(T)$ as determined from the two kind of
measurements coincide entirely. Interestingly, the line $H_{\rm
c2}(T)$ also presents an upward curvature. These data are fitted
nicely by the expression $H_{\rm c2}(T)=H_{\rm c2}(0)(1-T/T_c)^m$,
where $H_{\rm c2}(0)=31.8\pm 0.8$ kOe and $m=1.34\pm 0.03$.
Finally, we observe that the resistive transition is accomplished
at even higher fields $H_{\rm nor}^{\rm res}(T)$. The $H_{\rm
nor}^{\rm res}(T)$ line may mark the onset of surface
superconductivity \cite{James63,Abrikosov88}, as usually observed
in other low-$T_c$ superconductors \cite{Welp03,Shi03}. Transport
data and the concept of surface superconductivity will be
presented and discussed in detail for the Nb$-2$ film (see below).

One of the main results of the present paper, regarding the
subject of TMI, is shown in the low-field regime as presented in
detail in fig. \ref{b12}(b). We see that the $H_{\rm fp}(T)$,
$H_{\rm fj}(T)$ and $H_{\rm ''smp''}(T)$ lines connect at a
characteristic point $(H,T)\approx (80\ {\rm Oe}, 7.2\ {\rm K})$.
The $H_{\rm ''smp''}(T)$ line is placed in high fields, {\it while
the $H_{\rm fj}(T)$ line gradually moves in lower fields and below
$T\approx 6.4$ K takes the constant value $H_{\rm fj}(T<6.4 K)=40$
Oe.} This experimental fact, of a temperature independent $H_{\rm
fj}(T)$ line, is in contrast to theoretical predictions that treat
the $H_{\rm fj}(T)$ line as a simple boundary above which TMI
occur \cite{Swartz68,Wipf6791,Mints81,Mints96}. Indeed, within the
adiabatic approach and the Bean model
\cite{Swartz68,Wipf6791,Mints81}, for the case of bulk samples,
the first flux jump field $H_{\rm fj}$ exhibits a temperature
variation due to its dependence on the critical current $j_c(T)$
and on the specific heat $C(T)$, as
\begin{equation}
H_{\rm fj}(T)=\sqrt{\pi^3C(T)\frac{j_c(T)}{\mid\frac{dj_c(T)}{dT}\mid}}. \label{eq2}
\end{equation}
We may assume that in Nb, the specific heat in the superconducting
state is fairly described by $C(T)\approx C_oT^3$, and that the
non-linear temperature dependence $j_c(T)=j_o[1-(T/T_c)^n]^m$
holds for the critical current, in the most general case. The
result for the first flux jump field is
\begin{equation}
H_{\rm fj}(T)=H_oT^2\sqrt{(\frac{T_c}{T})^n-1}, \label{eq3}
\end{equation}
where $H_o=(\pi^3C_o/nm)^{1/2}$.

For the case of thin films, by rewriting Eq. \ref{eq1}, the first
flux jump field is given by \cite{Mints96}
\begin{equation}
H_{\rm fj}(T)=\frac{1}{\mu
_0\acute{H}l}\frac{\kappa(T)q^2j_c(T)}{\mid
\frac{dj_c(T)}{dT}\mid}. \label{eq4}
\end{equation}
For Nb the thermal conductivity is described by the relation
$\kappa(T)\approx \kappa _oT^\kappa$. In the superconducting state
($3$ K$<T<T_{c2}(H)$) the exponent $\kappa$ takes the values
$1<\kappa<3$ \cite{Ohara74Carlson70}. By assuming the general
relation $j_c(T)=j_o[1-(T/T_c)^n]^m$ for the critical current, we
arrive to the following equation
\begin{equation}
H_{\rm fj}(T)=H_1T^{\kappa+1}[({\frac{T_c}{T}})^{n}-1],
\label{eq5}
\end{equation}
where $H_1=(\kappa_o q^2/\mu_0 \acute{H}lnm)$.

A condition needed for flux jumps to occur is that the first flux
jump field $H_{\rm fj}(T)$ is placed below the full penetration
field \cite{Wipf6791,Mints81}, which may be experimentally
approximated by the first peak field $H_{\rm fp}(T)$. So, when
$H_{\rm fj}(T)<H_{\rm fp}(T)$ flux jumps are expected while above
a characteristic temperature $T_o$, where $H_{\rm
fj}(T>T_o)>H_{\rm fp}(T>T_o)$, flux jumps should not be observed.
For the case of a strip, the full penetration field is given by
\cite{Brandt96}
\begin{equation}
H_{\rm fp}(T)=\frac{d}{\pi}[1+ln(\frac{w}{d})]j_c(T), \label{eq6}
\end{equation}
where $w$ and $d$ are the width and the thickness
of the film respectively, and $j_c(T)=j_o[1-(T/T_c)^n]^m$ in the most general case.
So, the first peak field exhibits a behavior of the form
\begin{equation}
H_{\rm fp}(T)=H_2[1-(\frac{T}{T_c})^n]^m, \label{eq7}
\end{equation}
with $H_2=(d/\pi)[1+ln(w/d)]j_o$.

\begin{figure}[tbp] \centering%
\includegraphics[angle=0,width=8cm]{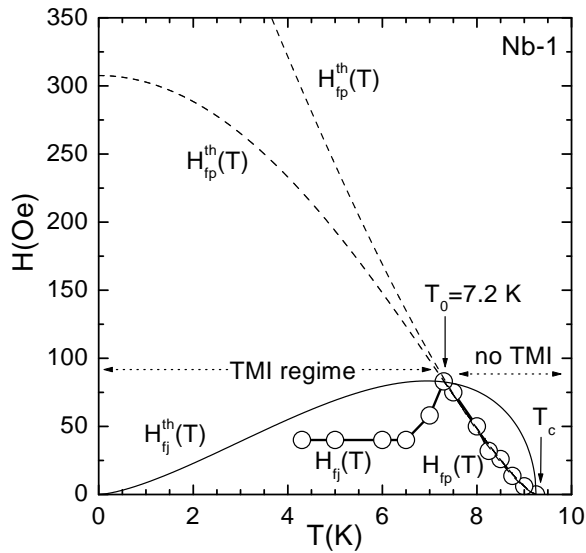}%
\caption { Experimental data (circles) and theoretical fitting
curves by using Eq. \ref{eq7} (dot lines) for the first peak field
$H_{\rm fp}(T)$ for $T>T_o$. In the temperature regime $T<T_o$ the
data refer to the first flux jump field $H_{\rm fj}(T)$, while the
solid curve reproduces the theoretical expression of Eq. \ref{eq3}
with the parameters $H_o=3$ and $n=1$. We observe the inability of
the theoretical curve (solid line) to describe the experimental
data (see text for more details). }
\label{b13}%
\end{figure}%

In fig. \ref{b13} we present the comparison of our experimental
data to the outlined theoretical suggestions. We clearly see that
in the temperature range $T>T_o=7.2$ K, where the line $H_{\rm
fj}(T)$ tends to overcome the first peak line $H_{\rm fp}(T)$, TMI
are no longer observed. In this temperature regime ($T>T_o=7.2$
K), the first peak field $H_{\rm fp}(T)$ is experimentally well
defined, and is described nicely by the theoretical expression of
Eq. \ref{eq7}. The dotted curve, placed at high fields, refers to
the fitting parameters $n=1$, $m=1.34$ and $H_2=686\pm 11$ Oe,
while the dotted one which is placed in lower fields, is described
by $n=2$, $m=1.34$ and $H_2=308\pm 5$ Oe. The different exponents
$n=1$ and $2$ refer to the most common cases discussed in the
literature, where the critical current exhibits a linear and a
quadratic dependence on $T/T_c$. We must underline that in both
cases the critical temperature was treated as a constant, equal to
the real $T_c=9.25$ K, and the exponent $m=1.34$ has the same
value as in the case of the fitting procedure for the
upper-critical field $H_{\rm c2}(T)$ (see fig. \ref{b12}(a)). In
contrast, regarding $H_{\rm fj}(T)$ in the regime $T<T_o=7.2$ K
the experimental data are not described by any theoretical curve.
The solid line reproduces the theoretical curve of Eq. \ref{eq3}
for the parameters $H_o=3$ and $n=1$. This expression holds for
the case of bulk samples. The relation \ref{eq5}, that refers to
thin films, also fails to reproduce our experimental data,
regardless of the choice for the exponents $\kappa$ and $n$.

As we noticed above, the main discrepancy between current
theoretical suggestions and our experimental results, is the low
temperature behavior of the first flux jump line $H_{\rm fj}(T)$.
Our data clearly show that for $T<6.4$ K the $H_{\rm fj}(T)$ line
is actually constant, $H_{\rm fj}(T<6.4 K)=40$ Oe. Let us assume
that Eq. \ref{eq4} (referring to a thin film sample), or even Eq.
\ref{eq2} (referring to bulk samples) should be used to describe
the low-temperature part of our $H_{\rm fj}(T)$ line. The left
side of these equations should then be treated as a constant. By
solving these simple differential equations, we easily see that a
constant first flux jump line $H_{\rm fj}(T)$ requires an
exponential temperature decrease of the critical current
$j_c(T)=C_1exp(-C_2T^p)$ ($p>0$). This is at odds to well known
theoretical results or experimental findings that deal with the
pinning mechanism of vortices and the temperature variation of the
critical current $j_c(T)$ in low-$T_c$ superconductors. More
theoretical work is needed in order to resolve this discrepancy.

\subsection{Comparison with magneto-optical studies}

Let us now compare our results to recent magneto-optical studies.
Such studies performed in Nb films of similar thickness ($5000$
{\AA}), dimensions ($3\times 8$ mm$^2$) and quality ($T_c=9.1$ K)
as our films, revealed that below the reduced temperature
$T/T_c\approx 0.65$, the penetration of magnetic flux takes place
in the form of dendrites \cite{Duran95}. In our case the ''SMP''
line $H_{\rm ''smp''}(T)$ ends at $T_o\approx 7.2$ K, which in
reduced temperature units is $T_o/T_c\approx 0.78$. Duran et al.
showed that once a dendrite is formed, it remains {\it ''frozen in
place''} until the magnetic field is changed sufficiently, so that
a new structure to appear \cite{Duran95}. In our Nb films we
observed that when having waited for $5$ min before measuring, the
structure of the resulted loops remained exactly the same (with
slightly lower values due to the relaxation of the magnetization,
see fig. \ref{b8}(a)). This probably indicates a {\it
''stationary''} character of the vortex penetration process. This
result may have a common origin with the dendrites that were
observed in Ref. \onlinecite{Duran95} to remain {\it ''frozen in
place''}.

For the case of thin films of the recently discovered
superconductor MgB$_2$ it was proved that such dendritic
instabilities, as the ones observed by Duran et al. in Nb,
\cite{Duran95} are directly related to the flux jumps occurring in
magnetization measurements
\cite{Barkov03,Johansen02,Johansen01,Zhao02}. This has also been
confirmed recently in Nb$_3$Sn films \cite{Rudnev03}. On the other
hand, the magnetic studies of Refs.
\onlinecite{Barkov03,Johansen02,Johansen01,Zhao02} and
\onlinecite{Rudnev03} didn't exhibit a ''SMP''. Despite that, the
overall similarity of their observations on TMI with our results,
prompts us to assume that even for the case of Nb films the flux
jumps, and consequently the formation of the ''SMP'', are probably
related to dendrites of vortices. Recent numerical simulations on
magnetic flux penetration in type-II superconductors also advocate
to this point of view \cite{Aranson01}. It would be interesting if
magneto-optical studies could reveal new information about the
possible relation of dendritic structures and the ''SMP'' observed
in our Nb$-1$ film.

\section{Experimental results and discussion for sample Nb$-2$}

\subsection{Magnetoresistance measurements and the peak effect}

\begin{figure}[tbp] \centering%
\includegraphics[angle=0,width=8cm]{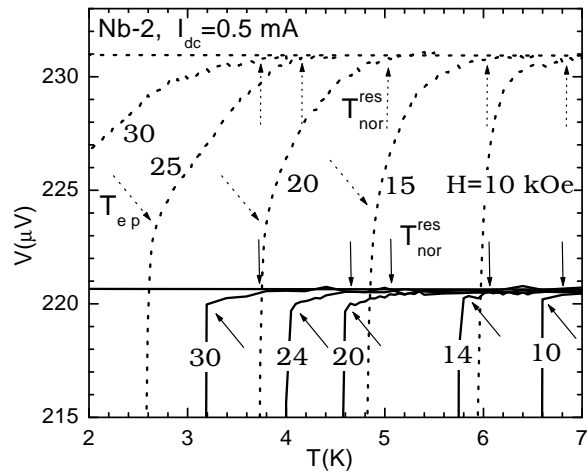}%
\caption { Measured voltage as a function of temperature for film
Nb$-2$, under various magnetic fields when normal (dot curves) and
when parallel (solid curves) to the surface of the film. In both
cases the applied current ($I_{\rm dc}=0.5$ mA) was normal to the
magnetic field ${\bf J}_{\rm dc}\perp{\bf H}_{\rm dc}$. Inclined
arrows denote the end points $T_{\rm e\ p}$ of the voltage drop,
while vertical arrows refer to the points $T_{\rm nor}^{\rm res}$
where the voltage takes the normal state value. }
\label{b1}%
\end{figure}

The situation  is different for the Nb$-2$ film, which is thinner
(1600 {\AA}) and is produced without annealing during the
deposition. Figure \ref{b1} presents two sets of magnetoresistance
measurements, performed at different directions of the external
magnetic field. At the first set (dotted curves) the dc field was
normal, while at the second set (solid curves) was parallel to the
surface of the film. In both sets of data we observed a sharp
decrease of the voltage at the end points $T_{\rm e\ p}$, which
are denoted by the inclined arrows. Above $T_{\rm e\ p}$, the
voltage curves gradually take the normal state value at the
characteristic points $T_{\rm nor}^{\rm res}$. Furthermore, we see
that the sharp drop of the voltage at points $T_{\rm e\ p}$
becomes more evident when high magnetic fields are applied.
Interestingly, the situation is analogous to the results observed
by magnetotransport measurements for the case of the cubic
(K,Ba)BiO$_3$ superconductor in Ref. \onlinecite{Blanchard02}. In
our case, as in that study, the voltage takes about $95$\% of its
normal state value at $T_{\rm e\ p}$, and after increases up to
$100$\% at $T_{\rm nor}^{\rm res}$.

\begin{figure}[tbp] \centering%
\includegraphics[angle=0,width=8cm]{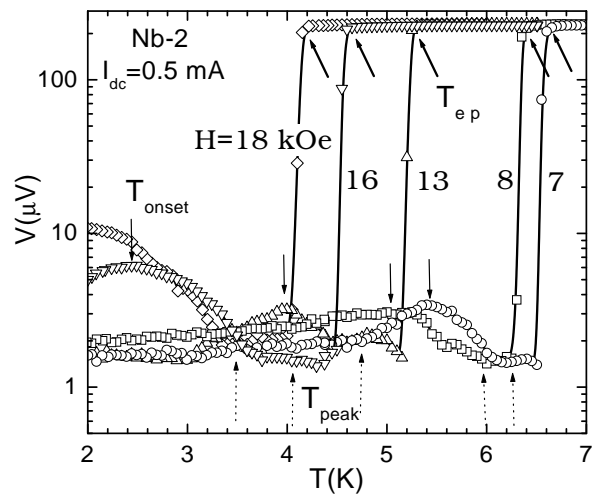}%
\caption { Semi-logarithmic plot of the measured voltage as a
function of the temperature for film Nb$-2$, under a dc transport
current $I_{{\rm dc}}=0.5$ mA, for various dc magnetic fields
$H_{{\rm dc}}=7, 8, 13, 16$ and $18$ kOe (${\bf H}\parallel{\bf
c}$). We observe three characteristic points: (a) $T_{\rm onset}$
where a peak is formed, (b) $T_{\rm peak}$ where the local minimum
corresponds to a peak in the critical current and (c) $T_{\rm e\
p}$ where a sharp increase occurs. }
\label{b2}%
\end{figure}

Careful measurements, performed in the whole temperature regime of
the mixed state, in the Nb$-2$ film, revealed that for $T<T_{\rm
e\ p}$ the signal is not really zero (as happened in film Nb$-1$)
but possesses structure. In fig. \ref{b2} we present, in a
semi-logarithmic plot, representative measurements for the case
where the magnetic field is normal to the film's surface, (${\bf
H}\parallel{\bf c}$). We observe that, in addition to $T_{\rm e\
p}$, the measured voltage curves exhibit two more characteristic
points. First, a local peak at $T_{\rm onset}$, and second a dip
at $T_{\rm peak}$. This is the well-known PE in the critical
current $J_c$ (the minimum in the measured voltage corresponds to
a maximum/peak in the critical current $J_c$), which was observed
in high-$T_c$ compounds
\cite{Kwok92,Sarkar01,Stamopoulos01,Stamopoulos02}, in
superconductors of intermediate-$T_c$ as pristine and Carbon-doped
MgB$_2$ \cite{PissasPRL,Welp03,PissasNEW}, in Nb
\cite{Autler62,Sorbo64,Tedmon65,Gammel98,Ling01} and other
low-$T_c$ \cite{Tomy97,Ling98,Banerjee98,Sarkar01} disordered
superconductors. The comparison to our magnetic measurements (not
shown here) revealed that the PE almost coincides with the
upper-critical fields (see fig. \ref{b4} below). As we move in the
high-temperature regime the PE is reduced.

\begin{figure}[tbp] \centering%
\includegraphics[angle=0,width=8cm]{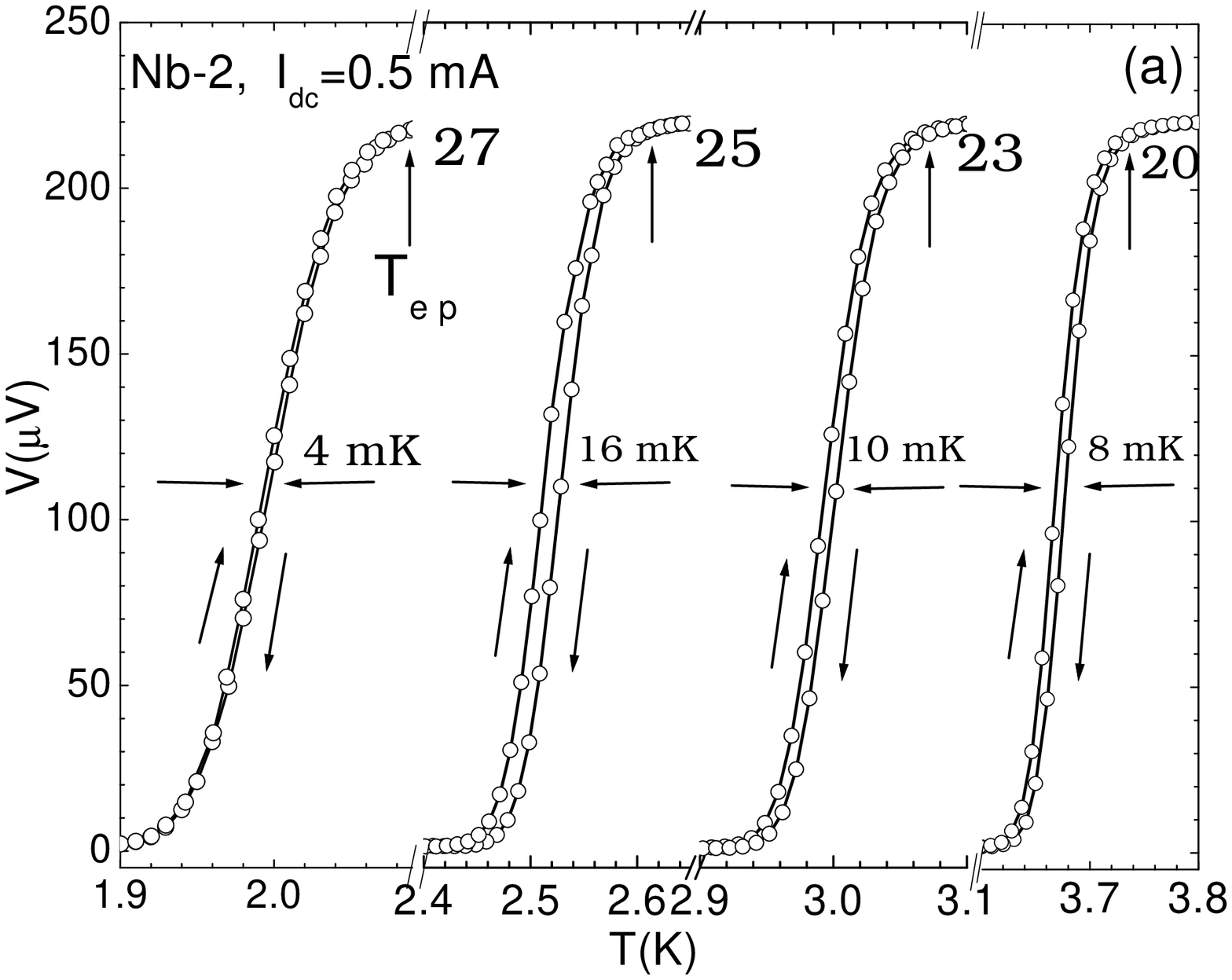}
\includegraphics[angle=0,width=8.5cm]{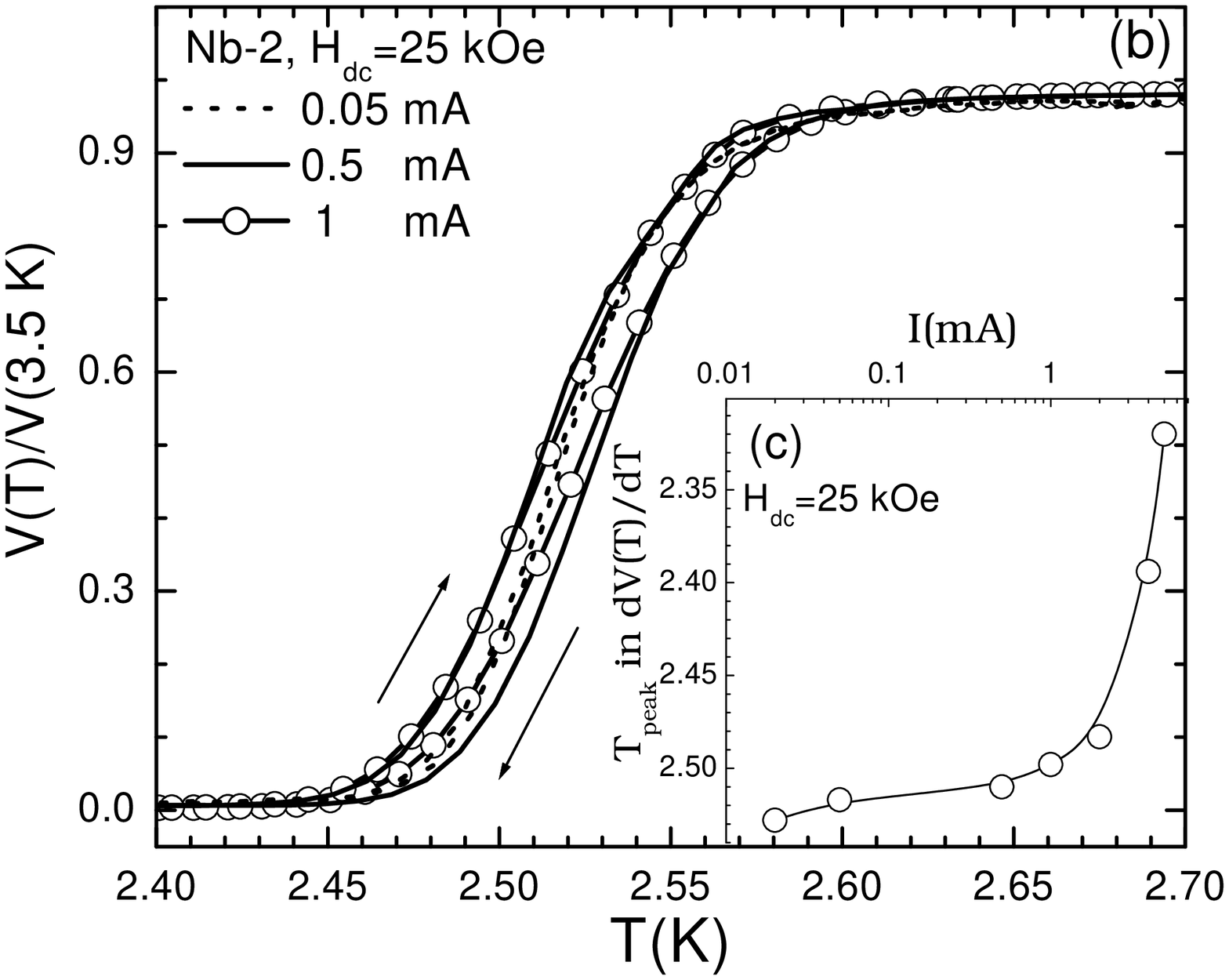}%
\caption { Measured voltage while increasing and decreasing the
temperature for Nb$-2$ (a) under a dc transport current $I_{{\rm
dc}}=0.5$ mA, for various dc magnetic-fields $H_{{\rm dc}}=20, 23,
25$ and $27$ kOe and (b) under a dc magnetic-field $H_{{\rm
dc}}=25$ kOe for various dc transport currents $I_{{\rm dc}}=0.05,
0.5$ and $1$ mA (the curve corresponding to $I_{{\rm dc}}=1$ mA is
shifted for clarity). In the inset (c) we present the variation of
the peak in the derivative $dV(T)/dT$ as a function of the applied
dc current. In all cases ${\bf H}\parallel{\bf c}$. }
\label{b3}%
\end{figure}

A question which remains open is the nature of the PE and of the
residual resistive transition in the temperature regime $T_{\rm
peak}<T<T_{\rm nor}^{\rm res}$. To investigate the underlying
physical processes in detail, we performed measurements for
increasing and decreasing the temperature. In fig. \ref{b3}(a) we
present the results for a constant current, in various dc fields.
We observed that in the temperature interval $T_{\rm
peak}<T<T_{\rm e\  p}$ the response is hysteretic. The hysteresis
is very narrow, $\Delta T<20$ mK, and is suppressed in the high
and low-field regimes, presenting a maximum for intermediate
values of the applied magnetic-field. The influence of the
transport current on the hysteretic response was also investigated
for currents $0.02$ mA$<I_{\rm dc}<5$ mA. Representative results,
for the case where $H_{\rm dc}=25$ kOe are summarized in fig.
\ref{b3}(b). For very small transport currents the hysteresis is
totally suppressed, as this is evident for the case where $I_{\rm
dc}=0.05$ mA. When the applied current is of intermediate values,
hysteretic behavior is clearly detected, as we present for the
case where $I_{\rm dc}=0.5$ mA. For even higher currents the
hysteresis is reduced, as this is evident when $I_{\rm dc}=1$ mA.
At the end, for $I_{{\rm dc}}>5$ mA the hysteresis was totally
suppressed (curves not shown). We carefully checked the
reproducibility of our results. Temperature steps were limited to
$10$ mK in most of our measurements. In order to ensure thermal
homogeneity in the whole film, the temperature sweep was very
slow, approximately $6$ mK/min. To improve the signal-to-noise
ratio, at every temperature we collected data for almost one
minute. The observed hysteretic behavior could not be caused by a
limited temperature resolution or stabilization inability of our
SQUID. Successive measurements, performed under the same
experimental conditions, revealed that the resulting curves $V(T)$
coincided within $\pm 2$ mK. In the inset we present the influence
of the applied current on the position of the peak as defined from
the derivative $dV(T)/dT$. When small currents are applied, the
position of the derivative's peak is almost constant but above a
threshold value ($I_{\rm dc}\approx 1$ mA) it becomes strongly
current dependent. These experimental results for the Nb$-2$ film
are discussed below, where the phase diagram of vortex matter is
introduced.

\subsection{Phase diagram of vortex matter for Nb$-2$.
Comparison with theory and other experimental works}

\begin{figure}[tbp] \centering%
\includegraphics[angle=0,width=8cm]{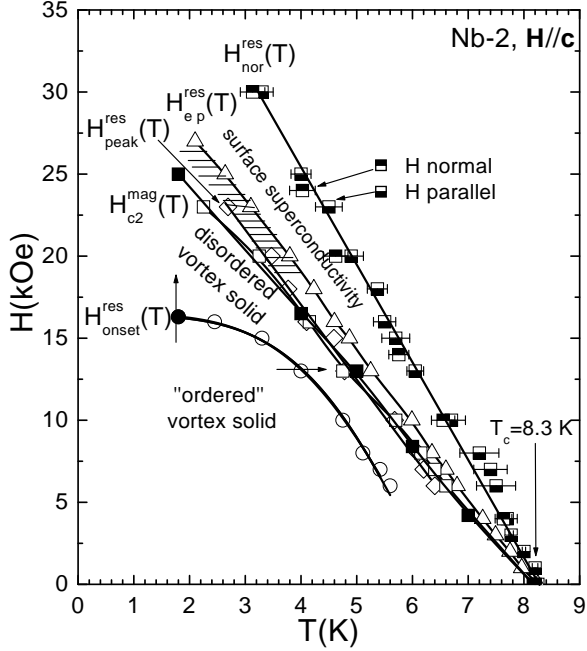}
\caption { Phase diagram of vortex matter for the Nb$-2$ film
(${\bf H}\parallel{\bf c}$). Presented are the peak's onset
$H_{\rm onset}^{\rm res}(T)$ (open and solid circles coming from
$V(T)$ and $V(H)$ measurements respectively), the peak $H_{\rm
peak}^{\rm res}(T)$ (rhombi), the end points of the peak $H_{\rm
e\  p}^{\rm res}(T)$ (triangles) and the upper-critical field
$H_{\rm c2}^{\rm mag}(T)$ (solid and open squares coming from m(T)
and m(H) magnetic measurements respectively), for the case where
the magnetic field is normal to the film's surface. Also presented
are the field lines $H_{\rm nor}^{\rm res}$ (semi-filled squares),
at where the voltage takes the normal state value, for both field
configurations. The shaded area indicates the regime where
hysteretic behavior has been observed in our $V(T)$ curves. }
\label{b4}%
\end{figure}

Our experimental results for the configuration where the field is
normal to the surface of the film are plotted in fig. \ref{b4}. In
the presented characteristic lines, the upper index {\it res}
({\it mag}) refers to data obtained by resistance (magnetic)
measurements. First of all, we see that the onset line $H_{\rm
onset}^{\rm res}(T)$ {\it saturates in the low temperature
regime}, and exhibits a monotonic decrease toward $T_c$. The
$H_{\rm peak}^{\rm res}(T)$, $H_{\rm e\ p}^{\rm res}(T)$ and
$H_{\rm c2}^{\rm mag}(T)$ lines maintain a linear temperature
dependence in the whole region investigated here. In the regime
close to $T_c$, the onset and the PE lines are strongly
suppressed. Maybe this is not the actual behavior of vortex matter
but it is caused by the limited resolution of our voltmeter.
Despite that, the same behavior has been observed in other
low-$T_c$ superconductors
\cite{Sarkar01,Salem02,Tomy97,Banerjee98}.

The overall behavior presented in fig. \ref{b4} is also
qualitatively similar to the phase diagrams of vortex matter
observed in disordered high-$T_c$ superconductors, where close to
the critical temperature, the SMP and/or PE lines are placed very
close to the irreversibility line
\cite{Kupfer98,Blasius99,Giller97,Sarkar01,Stamopoulos01,Stamopoulos02,Stamopoulos03}.
A quantitative comparison to recently proposed theoretical models
could give further information for the underlying mechanisms that
motivate the observed similarity.

Let us start our discussion with the characteristic line $H_{\rm
onset}^{\rm res}(T)$ where the onset of the PE occurs. As we see
in Fig. \ref{b3} these data are placed well inside the mixed state
of the superconductor. Recently, it was proposed that an
order-disorder transition between two vortex solid states occurs
in point disordered superconductors
\cite{Ertas96,Vinokur98,Giller97}. Experimental studies confirmed
the order-disorder transition occurring at (or in close proximity
to) the onset of the SMP for disordered high-$T_c$
\cite{KhaykovichPRL,Giller99,Kokkaliaris00,Stamopoulos02,Giller97}
and at the onset of the PE for the case of low-$T_c$ \cite{Ling98}
superconductors. In the ''cage model'' of Refs.
\onlinecite{Ertas96} and \onlinecite{Vinokur98} a pinning
parameter $\gamma_p$ was introduced that characterizes the static
disorder. Depending on the particular pinning mechanism the
temperature variation of the pinning parameter is: $\gamma_p\sim
\lambda^{-4}$ for $\delta T_c$ pinning (variations in the local
transition temperature of the superconductor is the origin of
pinning) and $\gamma_p\sim (\xi\lambda)^{-4}$ for $\delta l$
pinning (variation of the electron mean free path act as pinning
centers) \cite{blatter94}. The transition field is estimated by
equating the pinning energy to the elastic energy of a vortex. The
final expression depends on the relation between three
characteristic lengths of the model: $L_{\rm o}\approx
2\varepsilon a_{\rm o}$ which is the length of the "cage" along
the $c-$axis, $L_c=(\varepsilon^4\varepsilon_{\rm
o}^2\xi^2/\gamma_p)^{1/3}$ which is the related pinning length of
the vortex and finally $d$ which is the interlayer distance
\cite{Vinokur98}. In these expressions $\varepsilon$ is the
anisotropy parameter ($\varepsilon =1$ in our case), $a_{\rm o}$
is the mean distance of vortices and $\varepsilon_{\rm o}$ is the
vortex line energy. By comparing the characteristic length scales
we distinguish three cases for the pinning energy: (i) If
$d<L_{\rm o}<L_c$ the pinning energy in the cage is $E_p\approx
(\gamma_p\xi^2L_{\rm o})^{1/2}\approx (2\gamma_p\varepsilon a_{\rm
o}\xi^{2})^{1/2}$. (ii) When $d<L_c<L_{\rm o}$ the pinning energy
becomes $E_p=E_{\rm dp}(L_{\rm o}/L_c)^{(2\zeta-1)}$ with
$2\zeta-1\approx 1/5$, where $E_{\rm
dp}=(\gamma_p\varepsilon^2\varepsilon_{\rm o}\xi^4)^{1/3}$ is the
depinning energy of a single vortex line. (iii) Finally, when
$L_c<d<L_{\rm o}$ we have 2D pinning and $E_p$ becomes,
$E_p\approx U_p(L_{\rm o}/d)^{1/5}$, where
$U_p\approx\pi\sqrt{\gamma_p\xi^2/d}$. The vortex lattice to
vortex glass transition field is estimated by equating the pinning
energy, $E_p$ to the elastic energy,
$E_e=\varepsilon\varepsilon_{\rm o}c_L^2a_{\rm o}$ of a vortex.
For each of the three cases mentioned above the transition field
becomes: (i) when $d<L_{\rm o}<L_c$ we have $H_{\rm
on}(T)\sim\varepsilon^2[1-(T/T_c)^p]^2$ for $\delta T_c$ pinning,
while $H_{\rm on}(T)\sim\varepsilon^2[1-(T/T_c)^p]^{-2}$ for
$\delta l$ pinning, (ii) when $d<L_c<L_{\rm o}$, $H_{\rm
on}(T)\sim\varepsilon[1-(T/T_c)^p]^{3/2}$ for $\delta T_c$
pinning, while $H_{\rm on}(T)\sim\varepsilon[1-(T/T_c)^p]^{-1/2}$
for $\delta l$ pinning. Finally, in case (iii) where $L_c<d<L_{\rm
o}$ (2D pinning regime), $H_{\rm
on}(T)\sim\varepsilon^2[1-(T/T_c)^p]^{5/4}$ for $\delta T_c$
pinning, and $H_{\rm on}(T)\sim\varepsilon^2[1-(T/T_c)^p]^{-5/4}$
for $\delta l$ pinning (for more details see Refs.
\onlinecite{Giller97,Ertas96,Vinokur98}). For isotropic Nb is
reasonable to assume that the conditions $L_{\rm o}>d$ and $L_c>d$
always hold. We thus fitted the onset points by the proposed
expression $H_{\rm on}(T)=H_o(1-(T/T_o)^p)^n$, which holds for
$\Delta T_c$ pinning mechanism. A least squares criterion yielded
$H_o=16.5$ kOe, $T_o=7$ K if we assume the exponents $p=4$ and
$n=2$ (case (i)), and $H_o=16.4$ kOe, $T_o=6.6$ K if we choose
$p=4$ and $n=3/2$ (case (ii)) \cite{Vinokur98}. These fitting
curves are denoted in Fig. \ref{b3} by two solid lines that
coincide in the entire temperature regime. We observe that the
equation proposed by the theory describes accurately our
experimental results.

The next characteristic line of the phase diagram is the
upper-critical field $H_{\rm c2}^{\rm mag}(T)$, which as we know
defines the points where the bulk of the superconductor enters the
normal state as we increase the temperature. As we observe, the
$H_{\rm c2}^{\rm mag}(T)$ line coincides entirely with the PE line
$H_{\rm peak}^{\rm res}(T)$. This has been observed not only in Nb
\cite{Autler62,Sorbo64,Tedmon65}, but also in MgB$_2$
\cite{Welp03} superconductor.

Let us now discuss, in more detail, the possible origin of the
$H_{\rm e\ p}^{\rm res}(T)$ and the $H_{\rm nor}^{\rm res}(T)$
lines. A brief comparison to other low or even high-$T_c$
superconductors is also made. The main reasons invoked to
interpret the irreversible magnetic behavior of isotropic
superconductors is the interplay of thermal fluctuations and
static disorder on vortex lines. In pure samples, thermal
fluctuations transforms the vortex lattice into a liquid of flux
lines through a first order transition. This has been observed in
the almost isotropic YBa$_2$Cu$_3$O$_{7-\delta }$
\cite{Charalambous92,Safar92,Kwok92}. On the other hand, in
disordered samples, it is the static disorder that transforms the
vortex lattice into an amorphous vortex solid. This behavior has
been observed in the high-$T_c$ compounds HgBa$_2$CuO$_{4+\delta
}$ \cite{Stamopoulos02} and YBa$_2$Cu$_3$O$_{7-\delta }$
\cite{Stamopoulos03} and in the low-$T_c$ superconductors
$2$H-NbSe$_2$ \cite{Ling98,Banerjee98}, CeRu$_2$ \cite{Banerjee98}
and X$_3$Rh$_4$Sn$_{13}$ \cite{Sarkar01,Tomy97} (X=Ca,Yb). In our
case, we observed a sharp drop ($\Delta T\approx 100$ mK) at the
end points of the PE which is also hysteretic. The observed
hysteresis is very narrow ($\Delta T< 20$ mK) and is restricted
exclusively in the regime $H_{\rm peak}^{\rm res}<H<H_{\rm e\
p}^{\rm res}$ (see shaded area in fig. \ref{b4}). These
experimental facts resemble the melting transition of vortex
matter. On the other hand, the sharp hysteretic drop, in the
measured voltage, occurs in the regime {\it above the magnetically
determined upper-critical field line $H_{\rm c2}^{\rm mag}(T)$},
placing beyond dispute that this effect doesn't refer to a
property of vortex matter. But one could object that our SQUID
magnetometer may underestimate the true upper-critical fields in
our thin Nb-$2$ film (thickness=$1600$ {\AA}), due to its limited
resolution. Even if this is true, there is a number of additional
arguments against a true solid-liquid phase transition. First, the
observed hysteresis faints when very low transport currents are
employed. This is not compatible to what is expected for a
transition of equilibrium origin, as was observed in YBCO
\cite{Kwok92,Charalambous92,Safar92}. Second, the decrease in the
measured voltage at the points $T_{\rm e\  p}$ is unconventionally
high, when compared to the resistive drop at the melting
transition in YBCO \cite{Kwok92,Charalambous92}. In addition, for
the case of the melting transition of vortex matter in YBCO, the
temperature sweeping down branch was placed above the sweeping up
one \cite{Kwok92,Charalambous92,Safar92}, while in our case we
observed the opposite behavior. These facts prevent us from
attributing the observed sharp features to the melting transition,
in contrast to previous studies in isotropic Nb films
\cite{Schmidt93} or even in MgB$_2$ single crystals
\cite{Pradhan02}.

It could be proposed, that the residual part of the resistive
transition in the regime $H_{\rm c2}^{\rm mag}(T)<H<H_{\rm
nor}^{\rm res}(T)$ (see fig. \ref{b4}), is motivated by the
mechanism of surface superconductivity. Then the sharp hysteretic
response, restricted in the regime $H_{\rm peak}^{\rm
res}(T)<H<H_{\rm e\ p}^{\rm res}(T)$ (see shaded area in fig.
\ref{b4}), refer to the phase transition at the bulk
upper-critical points, or is also motivated by surface
superconductivity. Below, let us briefly present the basic aspects
of surface superconductivity, so that a comparison with our
experimental results could be made. When physical boundary
conditions are taken into account for a magnetic field applied
parallel to the main surface of a superconductor, a new nucleation
field $H_{\rm c3}(T)$ exists, at where surface superconductivity
develops while lowering the temperature
\cite{James63,Abrikosov88}. This characteristic field is placed
above the true upper-critical field $H_{\rm c2}(T)$, and is
related to it through $H_{\rm c3}(T)=1.695H_{\rm c2}(T)$. For an
isotropic superconductor as Nb, the upper-critical field $H_{\rm
c2}(T)$ should be independent of the orientation of the applied
field. Thus, we should expect that, similarly, the surface
nucleation field $H_{\rm c3}(T)$ should occur at the same points
$1.695H_{\rm c2}(T)$, regardless of the orientation of the
magnetic field. As is evident in fig. \ref{b4} (see also fig.
\ref{b1}), the voltage takes the normal state value at almost the
same points $H_{\rm nor}^{\rm res}$ in both field configurations.
This could be an additional indication that, in both cases, the
experimental lines $H_{\rm nor}^{\rm res}(T)$ indeed mark the
surface superconductivity effect. At zero temperature, the ratio
of the surface superconductivity field $H_{\rm nor}^{\rm res}(0)$
to the upper-critical field $H_{\rm c2}^{\rm mag}(0)$ is $H_{\rm
nor}^{\rm res}(0)/H_{\rm c2}^{\rm mag}(0)\approx 1.45$. This is a
reasonable value if we keep in mind that the ratio $H_{\rm
nor}^{\rm res}(0)/H_{\rm c2}^{\rm mag}(0)$ depends on the
combination of two structural parameters that may vary from sample
to sample: first, the quality of the bulk (reflected in $H_{\rm
c2}^{\rm mag}(0)$) and second, the quality of the surface
(reflected in $H_{\rm nor}^{\rm res}(0)$). On the other hand, the
original theoretical treatment suggests that the surface
nucleation field $H_{\rm c3}(T)$ should not be observed when the
magnetic field is normal to the main surface of the film. In our
case, we observed the same qualitative behavior in both field
configurations. In the past, Drulis et al. \cite{Drulis91} did not
observe surface superconductivity in Nb foils for the case where
the applied field was normal to the surface of their samples, but
only for the parallel field configuration. We note that in Ref.
\onlinecite{Drulis91} only the regime of relatively low fields
($H<5$ kOe) was studied. Our results extend to much higher fields
($H\leq 30$ kOe). As is evident in figs. \ref{b1} and \ref{b4} the
effect under discussion is more pronounced in the high field
regime, where the lines $H_{\rm e\ p}^{\rm res}(T)$ and $H_{\rm
nor}^{\rm res}(T)$ diverge substantially. As a consequence, we
were able to detect surface superconductivity even for the normal
field configuration. Probably, the same behavior would have been
observed in Ref. \onlinecite{Drulis91}, had the measurements been
performed in higher applied fields. Therefore, we believe that our
results actually extend the currently available experimental data,
regarding the surface superconductivity in Nb. The same behavior
has been also observed, for the normal field configuration, in the
MgB$_2$ superconductor \cite{Welp03}. This discrepancy between
theory and experiment remains to be resolved.

Consequently, regarding the phase diagram of vortex matter for the
Nb-$2$ film, we believe that in the low-temperature-magnetic-field
regime a vortex quasi-lattice exists. Through the onset points
$H_{\rm onset}^{\rm res}(T)$ of the PE, a gradual disordering of
the vortex solid takes place as we approach the $H_{\rm c2}^{\rm
mag}(T)$. This is actually the softening effect of the vortex
lattice, as has been proposed by Larkin and Ovchinnikov many years
ago \cite{Larkin79}. At the end, above the line $H_{\rm c2}^{\rm
mag}(T)$ the effect of surface superconductivity is observed up to
$H_{\rm nor}^{\rm res}(T)$ .

\section{Crystallographic data and comparative discussion for the Nb$-1$ and  Nb$-2$ films}

In this last part of the present article we discuss
crystallographic results and we make a comparative discussion for
the two Nb films, in order to outline the possible mechanism that
motivate the observed similarities and/or differences. The
resistivity takes almost the same value $\varrho _n \approx 5-10$
$\mu \Omega$cm for both Nb films. In addition, the zero
temperature values of the coherence length $\xi(0)\approx 100$
{\AA}, the penetration depth $\lambda(0)\approx 900$ {\AA}, and
consequently the Ginzburg-Landau parameter $\kappa(0)\approx 9$,
are almost the same for the two Nb films \cite{kappa}. These
values are in fair agreement to the ones reported in the past in
disordered Nb samples
\cite{Drulis91,Schmidt93,Salem02,Esquinazi99}. Interestingly,
despite their almost identical values of $\xi(0)$, $\lambda(0)$
and $\kappa(0)$, the two films exhibit a completely different
behavior in the mixed state of their vortex phase diagrams. The
film Nb$-1$ (annealed during the deposition) presents the ''SMP''
and TMI in a great part of its phase diagram, while the film
Nb$-2$ (not annealed during the deposition) exhibits a
comparatively smooth PE, confined in the regime of the
upper-critical field $H_{\rm c2}(T)$.

We believe that the different behavior of vortex matter, in the
two kinds of films, is caused by the different preparation
conditions. Our combined x-ray diffraction (XRD) and transmission
electron microscopy (TEM) data revealed that by annealing the
films during deposition, a larger mean size of the grains is
produced, which in this case is $930$ {\AA} for Nb$-1$ (annealed)
and $420$ {\AA} for Nb$-2$ (not annealed). Furthermore, the grains
of the annealed film are oriented, in some degree, with $[110]$
direction perpendicular to the film's surface and exhibit a
tendency for columnar growth, while the film produced without
annealing doesn't show such a tendency. These crystallographic
data may give worthy information for the interpretation of the
mixed-state superconducting properties of the two different films.
It seems that in the mixed state of film Nb$-1$ the dynamic
behavior of vortices is governed by a correlated type of disorder
since its columnar growth could be considered as extended disorder
along the boundaries of the columns. This suggestion favorably
agrees with the results of Shantsev et al. \cite{Shantsev99}. In
that work \cite{Shantsev99} it was proved that a granular
structure of a superconducting specimen results in a peak in the
descending branch of the m(H) loop which is positioned in positive
field values. This behavior is observed in our Nb$-1$ film (see
upper inset of fig.\ref{b5}) and could be related to its tendency
for columnar growth. On the other hand in film Nb$-2$ the
mixed-state properties are governed mainly by point-like disorder.
As a result TMI are absent and only a smooth PE is observed.
Finally, the critical current density was estimated for both films
indirectly from magnetization loop and relaxation measurements
(data are not shown here since they are part of a subsequent
publication), and directly by measurements of the $I-V$
characteristics as the ones presented in fig.\ref{b11} for the
Nb$-1$ film. From those data the critical current density $J_c$
may be straightforwardly estimated for Nb$-1$ film. Since the
dimensions of the specific film are $0.3\times 0.3$ cm$^{2}$ and
its thickness is $7700$ {\AA} the effective current density is of
the order $J_{\rm dc}\simeq 0.1$ kA/cm$^{2}$. Thus, even in the
regime so close to the upper-critical field (for example at
$H=2.3$ kOe$\simeq H_{\rm c2}^{\rm mag}(8\rm K)$) the critical
current density $J_c$ is higher than $0.1$ kA/cm$^{2}$ (the
respective value at low temperatures exceeds $100$ kA/cm$^{2}$).
The respective critical current density of the Nb$-2$ film is
quite lower. All these facts mentioned above suggest that the
Nb$-1$ film is more disordered when compared to Nb$-2$.

We now discuss the superconducting properties of the films in the
regime near the superconducting-normal transition. Interestingly,
both films exhibit the same behavior in their residual resistive
transitions in the regime $H_{\rm c2}^{\rm mag}(T)<H<H_{\rm
nor}^{\rm res}(T)$. This effect could not be related to the bulk
properties of the two samples, because it occurs above the bulk
upper-critical field line $H_{\rm c2}^{\rm mag}(T)$. Furthermore,
as our XRD and TEM data revealed, the bulk structural properties
of the two films differ substantially. Thus, the detected effect
could not be related to the resistive properties of the bulk, but
probably to the superconducting properties of the surfaces. Thus
we conclude that the behavior of the resistive transition above
$H_{\rm c2}^{\rm mag}(T)$ is possibly related to surface
superconductivity.

\section{conclusions}

In summary, we presented magnetic and magnetotransport
measurements in films of the isotropic Nb superconductor. Film
Nb$-1$, prepared {\it under annealing} during the deposition,
exhibited TMI and a ''SMP'' feature in magnetic measurements. In
contrast to theoretical suggestions, the first flux jump field
$H_{\rm fj}$ is not inversely proportional to the sweep rate of
the applied field. TMI are observed up to the limiting temperature
$T_o=7.2$ K where the $H_{\rm fj}(T)$ and $H_{\rm fp}(T)$ lines
connect. Interestingly, in the low-temperature regime $T<6.4$ K,
the first flux jump field preserves a constant value $H_{\rm
fj}($T$<6.4$ K$)=40$ Oe. This is in strong disagreement to
theoretical proposals for thin film, or even bulk samples. The
comparison of our primary data obtained in film Nb$-1$ with
ancillary data obtained in a bulk sample of MgB$_2$, suggest that
the TMI exhibit noticeable differences when observed in films or
bulk samples. Our TEM and XRD data suggest that in Nb$-1$ the SMP
feature is probably motivated by the interaction of vortices with
some kind of correlated disorder existing probably due to the
tendency of films that are annealed during deposition to exhibit a
columnar growth.

On the other hand, the respective crystallographic data for the
less disordered Nb$-2$ film, which has not been annealed during
deposition, suggest that its mixed-state properties are governed
mainly by point-like disorder. As a result Nb$-2$ exhibited the
conventional PE in the vicinity of the upper-critical fields. The
end points of the PE exhibit a narrow hysteretic behavior ($\Delta
T< 20$ mK), when high magnetic fields are applied. Our results
indicate that sharp drops which are usually observed in
magnetoresistance data, obtained in low-$T_c$ superconductors,
should be interpreted with caution. Such findings should not be
directly related to a phase transition of vortex matter. It is
only measurements of equilibrium properties that can prove a true
phase transition of vortex matter. According to our results, the
mechanism of surface superconductivity is probably responsible for
the residual part of the magnetoresistance in the region $H_{\rm
c2}^{\rm mag}(T)<H<H_{\rm nor}^{\rm res}(T)$. By studying the high
field regime in detail, we showed that in contrast to current
theoretical treatment the effect is observed not only when the
magnetic field is parallel, but also when is normal to the surface
of our Nb films.

Finally, let us note that despite the different physical causes,
the resulting phase diagrams of vortex matter in our Nb disordered
films, have strong analogies to the ones observed in disordered
high-$T_c$ superconductors.

\begin{acknowledgments}
This work was supported by the IHP Network ''Quantum Magnetic
Dots'' Contract HPRN-CT-2000-00134 EU.
\end{acknowledgments}

\pagebreak


\begin{references}
\bibitem{dema} The magnetization is estimated when demagnetization corrections
are taken into account by claiming that the initial slope of the
measured curve should correspond to the Meissner state in which
$4\pi\mid M\mid$ equals the applied magnetic field. At the same
result we arrive if we correct the resulting magnetization by a
demagnetization factor $N$ estimated by the approximate formula
$N=1-(\pi/2)(d/R)$ which holds for the case of a flat disk of
thickness $d$ and ''diameter'' $R$, when $d\ll R$.\cite{Fetter67}
\bibitem{kappa} We estimate the Ginzburg-Landau parameter
$\kappa(0)=\lambda(0)/\xi(0)$ from the magnetization measurements.
The coherence length $\xi(0)\approx 100$ {\AA} is estimated from
the relation $H_{\rm c2}(T)={\Phi _o/2\pi \xi(T)^2}$ by
extrapolating the $H_{\rm c2}(T)$ curve down to zero temperature.
The zero-temperature penetration depth $\lambda(0)$ may be
estimated by the expression $\lambda(0)=\lambda _o\sqrt{\xi
_o/2.66l}$. The mean-free path, $l$ may be calculated from the
dirty-limit expression $\xi(0)\approx \sqrt{l\xi_{o}}$ which
results in $l\approx 23$ {\AA}, if we assume that $\xi _o=430$
{\AA}. \cite{Finnemore66} At the end, by assuming $\lambda _o=350$
{\AA} \cite{Finnemore66} we obtain $\lambda(0)\approx 900$ {\AA}.
The resulting Ginzburg-Landau parameter is $\kappa(0)\approx 9$.
\end{references}
\end{document}